\documentclass[aps,prl,showpacs,floats,twocolumn,floats,superscriptaddress,floatfix]{revtex4}
\usepackage{bm}
\usepackage{natbib}
\usepackage{amssymb}
\usepackage{amsfonts}
\usepackage{amsmath}
\usepackage{color}
\usepackage{times}
\usepackage{verbatim}
\usepackage{graphicx}
\usepackage{graphics,epsfig}
\usepackage{theorem}
\usepackage{makeidx}
\usepackage{amsmath}
\usepackage{epic}\usepackage{amscd}
\usepackage{bbm}
\usepackage{xy}
\usepackage{amssymb}
\usepackage{epsfig}

\newcommand{\ud}{\,\mathrm{d}}

\def\gsim{\;\rlap{\lower 2.5pt
\hbox{$\sim$}}\raise 1.5pt\hbox{$>$}\;}
\def\lsim{\;\rlap{\lower 2.5pt
\hbox{$\sim$}}\raise 1.5pt\hbox{$<$}\;}
\begin{document}

\newif\iffigs 
\figstrue
\iffigs \fi
\def\drawing #1 #2 #3 {
\begin{center}
\setlength{\unitlength}{1mm}
\begin{picture}(#1,#2)(0,0)
\put(0,0){\framebox(#1,#2){#3}}
\end{picture}
\end{center} }

\title{Excess Clustering on Large Scales in the MegaZ DR7 Photometric Redshift Survey}
\author{Shaun A. Thomas}
\author{Filipe B. Abdalla}
\author{Ofer Lahav}
\affiliation{Department of Physics and Astronomy, University College London, Gower Street, London, WC1E 6BT, UK}  

\begin{abstract}
We observe a large excess of power in the statistical clustering of Luminous Red Galaxies in the photometric SDSS galaxy sample called MegaZ DR7. This is seen over the lowest multipoles in the angular power spectra $C_{\ell}$ in four equally spaced redshift bins between $0.45 \le z \le 0.65$. However, it is most prominent in the highest redshift band at $\sim 4\sigma$ and it emerges at an effective scale $k \lesssim 0.01 \mathrm{h \, Mpc^{-1}}$. Given that MegaZ DR7 is the largest cosmic volume galaxy survey to date (3.3 (Gpc $h^{-1}$)$^3$) this implies an anomaly on the largest physical scales probed by galaxies. Alternatively, this signature could be a consequence of it appearing at the most systematically susceptible redshift. There are several explanations for this excess power that range from systematics to new physics. This could have important consequences for the next generation of galaxy surveys or the $\Lambda$CDM model. We test the survey, data and excess power, as well as possible origins.
\end{abstract}

\maketitle

{\it Introduction} -- A galaxy survey contains vast and varied information related to cosmological physics. The galaxies act as tracers of the underlying mass distribution whose statistical clustering enables a determination of the cosmological model. This is complementary to the CMB and directly probes the late-time Universe. Historically, this statistical distribution has been a penetrating indicator of new physics: The analysis of the galaxy correlation function in the APM survey showed one of the first signs that $\Omega_{m} < 1$ \citep{Maddox90,Efstathiou90} -- before supernovae. 

Since then galaxy surveys have matured with an emphasis on large cosmic volumes, which probe new scales, and immense galaxy numbers. Due to limited resources one approach has been to compensate this demand with a decrease in redshift precision. Rather than spectroscopy one instead obtains a redshift estimate based on the overall flux through broad band filters. This is called a photometric redshift. It has resulted in the leading MegaZ DR7 photometric catalogue \citep{Thomas10} and is the basis for the Dark Energy Survey \citep{DES05}. 

Any deviations from the current $\Lambda$CDM statistical profile could be the sign of new physics emerging over the largest scales in the Universe. Alternatively, any signatures could be systematics that affect the photometric method and therefore future projects. Detecting these systematics is vital in not only avoiding a biased inferred cosmology but in the planning of surveys.  

In this {\it letter} we highlight anomalous signatures in the {\it new} MegaZ DR7 galaxy angular power spectra given by excess power over large scales probed in the late-time Universe. We test the survey, excess power, possible systematics and highlight a subset of potential theoretical explanations.

{\it MegaZ} -- The MegaZ DR7 survey \citep{Thomas10} is a new catalogue of Luminous Red Galaxies (LRGs) based on the final Sloan Digital Sky Survey (SDSS) II photometric release \citep{Abazajian09}. It is an update of the previous DR4 catalogue given by \citep{Collister07,Blake07}. Covering almost one fifth of the entire sky it includes $723,556$ LRGs in four equally spaced redshift bins with width $\Delta z = 0.05$ between $0.45 < z < 0.65$. 

We performed a spherical harmonic analysis of the galaxy distribution \citep{Thomas10} by determining the theoretical angular power spectra,

\begin{equation} \label{eq:angularpowerspectrum}
C^{ij}_{\ell} \equiv <\delta^{2D} \delta^{* 2D}> = 4\pi \int \Delta^{2}(k) W_{i}(k) W_{j}(k) \frac{\ud k}{k}
\end{equation}
\noindent
for all four bins, where $\Delta^{2}(k)$ is the dimensionless power spectrum. This is the projected power spectrum with window functions given by $W_{i}(k)$ for bin $i$ under consideration. This is further detailed by $W_{l}(k) = \int f(z)j_{l}(kz) \ud z$ and $f(z) = n(z) D(z) (\frac{\ud x}{\ud z})$, with the spherical Bessel function $j_{l}(kz)$, the linear growth factor $D(z)$, comoving coordinate $x$ and the normalised redshift distribution $n(z)$. Redshift space distortions act to alter the shape of $C_{\ell}$ and we include this as described in \citep{Thomas10,Fisher94,Padmanabhan07}. The galaxies' photometric redshifts are determined with the ANNz code \citep{Collister04}, using the spectroscopically {\it and} photometrically defined 2SLAQ survey \citep{Cannon06} as a representative training set as in \citep{Abdalla08}. This was found to give the best photometric redshift estimate compared to template based methods \citep{Thomas10,Abdalla08}. The LRGs provide reliable photometric redshifts given that they are old, red, elliptical systems with a stable spectral energy distribution and sharp $4000 \AA$ break. Furthermore, due to their high luminosity, they probe a vast region of cosmic volume and therefore the largest of physical scales. The measured $C_{\ell}$ are determined using,
\begin{equation} \label{eq:incompleteskyCl}
C_{l,m}^{\mathrm{psky}} = \frac{|a_{l,m} - \frac{N}{\Delta \Omega}I_{l,m}|^{2} }{J_{l,m}} - \frac{\Delta \Omega}{N}
\end{equation}
\noindent
where $a_{l,m}$ are the harmonic coefficients of the projected distribution and $\Delta \Omega$ and $N$ correspond to survey area and number of galaxies, respectively. The coefficient corrections are given as,
\begin{equation} \label{eq:ilm}
I_{l,m} = \int_{\Delta \Omega} Y^{*}_{l,m} \ud\Omega \qquad J_{l,m} = \int_{\Delta \Omega} |Y_{l,m}|^{2} \ud\Omega.
\end{equation}
\noindent
Due to statistical isotropy the $C_{l,m}$ are averaged over $(2\ell + 1)$ $a_{l,m}$ values. These points are further binned into multipole bands of width $\Delta \ell = 10$ to decorrelate the data; a consequence of the partial sky coverage and subsequent convolution. Further details can be found in \citep{Peebles73,Blake07,Thomas10}, in addition to the full DR7 data in \citep{Thomas10}. Moreover, the constraining potential of this powerful data set has been shown for neutrino masses in \citep{Thomas09b}.

\begin{figure}
  \includegraphics[width=8.4cm,height=8.2cm]{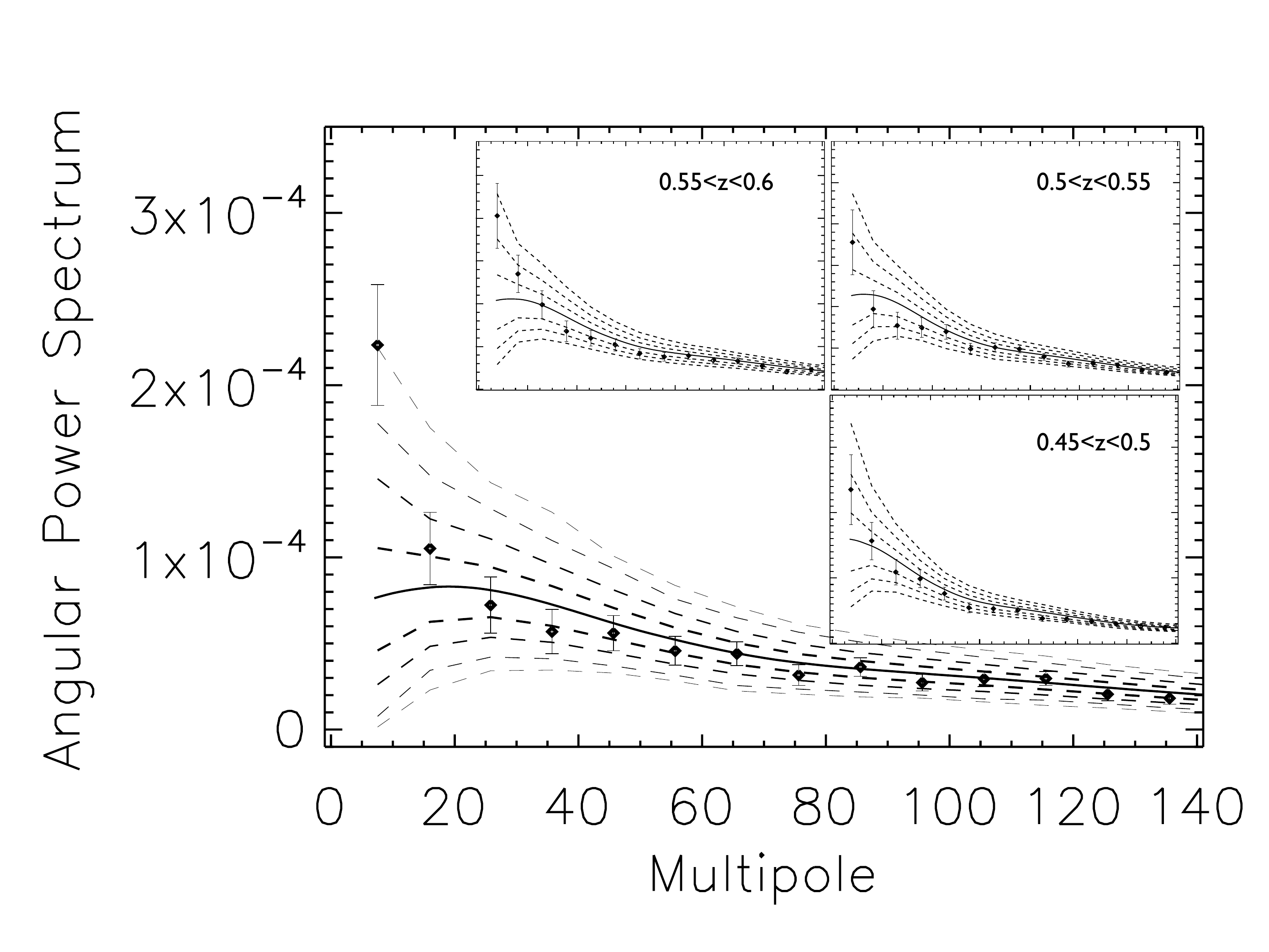}
    \caption{\small{The Angular Power Spectra $C_{\ell}$ measured in the SDSS photometric MegaZ DR7 Luminous Red Galaxy survey. The panels relate to four redshift bins with width $\Delta z = 0.05$ from $z = 0.45$ to $z=0.65$. The best fit theoretical spectra (solid lines) are excellent matches to the data including multipoles up to $\ell \sim 500$. However, the largest angular scales are observed to be anomalous; the dashed lines correspond to $1 \to 4$ $\sigma$ derived from simulations. This is particularly severe in the highest redshift bin (main panel), which is $\sim 4\sigma$.}}
    \label{fig:neutrinobestfitdata}
\end{figure}

{\it Excess Power} -- The measured $C_{\ell}$ for the four redshift bins are shown in Figure~\ref{fig:neutrinobestfitdata}. This illustrates the excess power over the lowest multipole bands compared to the rest of the data and the best fit theoretical profile (solid line). It is particularly prominent in the highest redshift bin ($0.6<z<0.65$; main panel). The model error bars in this plot have been assigned using Equation~\ref{eq:gaussianerror}. This accounts for the expected shot noise, survey area and cosmic variance.

\begin{equation} \label{eq:gaussianerror}
\sigma(C_{l}) = \sqrt{\frac{2}{f_{\mathrm{sky}}(2l+1) } } \Big( C_{l} + \frac{\Delta \Omega}{N} \Big)
\end{equation}
\noindent
In this way the excess discrepancies in each bin are seen to be severe. In order to more accurately quantify the anomalies we reconstruct $3 \times 10^{4}$ Gaussian realisations of the galaxy field from the best fit $C_{\ell}$s. We account for cross correlations between the redshift bins due to photometric uncertainty and impose the same DR7 mask, pipeline and measured galaxy number on the simulations. We successfully reconstructed the smooth input cosmology from these realisations for $\ell \geq 4$ (our cut due to the partial sky) and we also reproduced the previous MegaZ results. This demonstrated the reliability of the statistic in relation to the survey geometry and the treatment of shot noise. We then used the variance of these realisations to obtain a measure of the excess power uncertainty. The corresponding significance levels ($1 \to 4 \sigma$) are included as the dashed lines in Figure~\ref{fig:neutrinobestfitdata}. Just two of the realisations were as anomalous as the measured value in the lowest multipole band in the highest redshift bin. This gives a significance of $\sim 4\sigma$. Similarly, the lowest bands in the other bins are discrepant by $\sim 2\sigma$ (bin 1), $\sim 2\sigma$ (bin 2) and $\sim 2.5\sigma$ (bin 3). With slightly less significance $\sim 1\sigma$ excesses are observed for the second multipole bands in bin 3 and bin 4 too. The effect from including or excluding the excess power in the inferred cosmological constraints can be sizeable as shown in Figures 8 and 11 from \citep{Thomas10}. It is interesting that slight hints of excess power have also been alluded to in \citep{Blake07,Padmanabhan07,Huetsi09} with $P(k)$ and $C_{\ell}$ statistics.  

\begin{figure} 
    \centering
      \includegraphics[width=8cm,height=5.5cm]{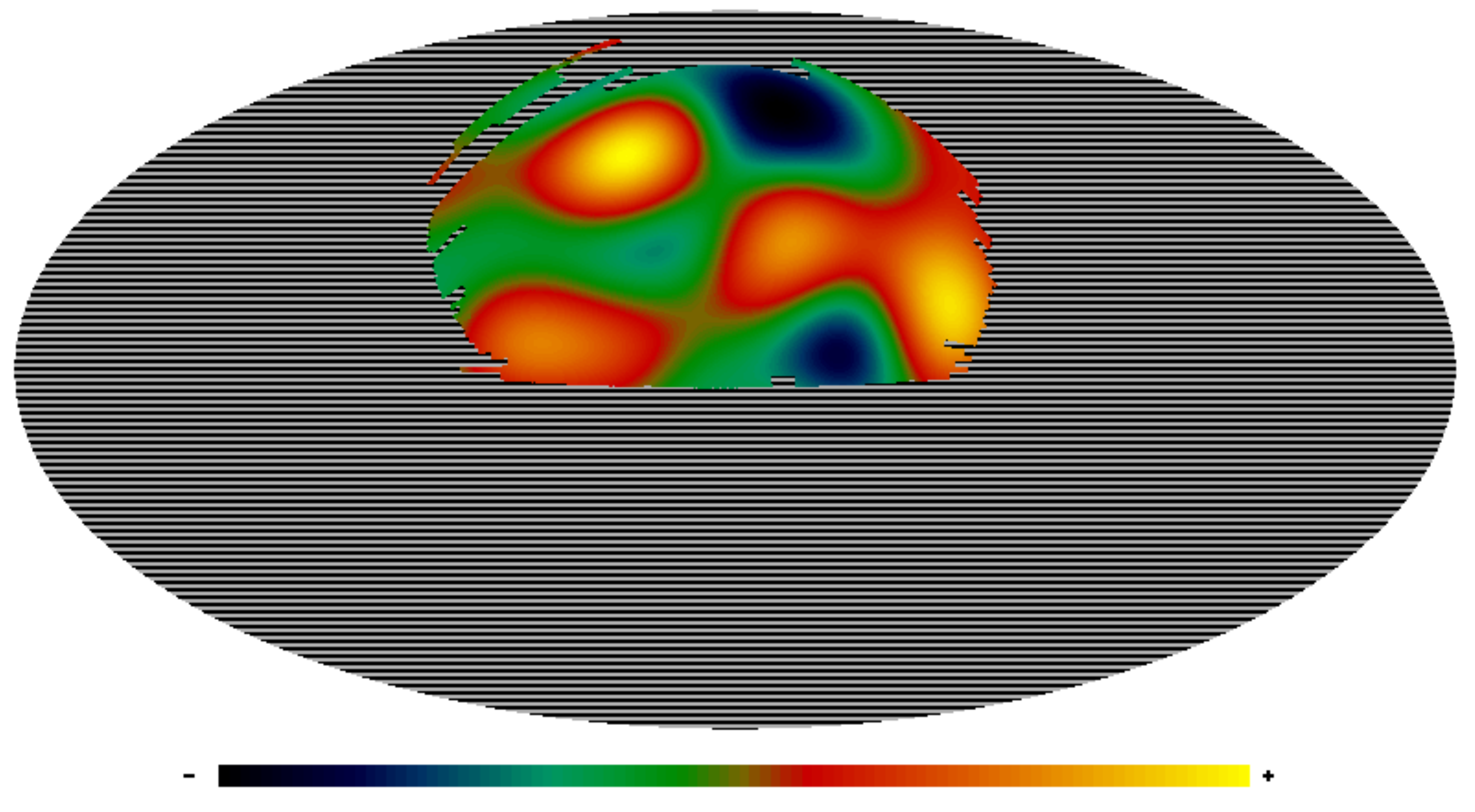}
    \caption{\small{A visualisation of the measured underlying field within the surveyed region on the sky. Only contributions from $\ell = 4 \to 10$ in the most anomalous multipole and redshift band are included.  }}
    \label{fig:almvisualisation}
\end{figure}

To assess potential contributions to the most anomalous of the $\ell$ bands we reconstruct the underlying matter distribution implied by that data using {\sc HEALPix} \citep{Gorski05}. The corresponding $a_{l,m}$ $\to$ matter visualisation can be observed across the surveyed region in Figure~\ref{fig:almvisualisation}. No clear pattern can be seen that would suggest the presence of an obvious systematic across the sky. I.e., there does not seem to be spurious contributions towards the edges of the survey, closer to the Galactic plane or concentrated in one surveyed region. However, we now strive to quantify candidate sources of contamination.

\begin{figure*}
  \begin{flushleft}
    \centering
    \begin{minipage}[c]{1.00\textwidth}
      \centering
      \includegraphics[width=8.4cm,height=8.4cm]{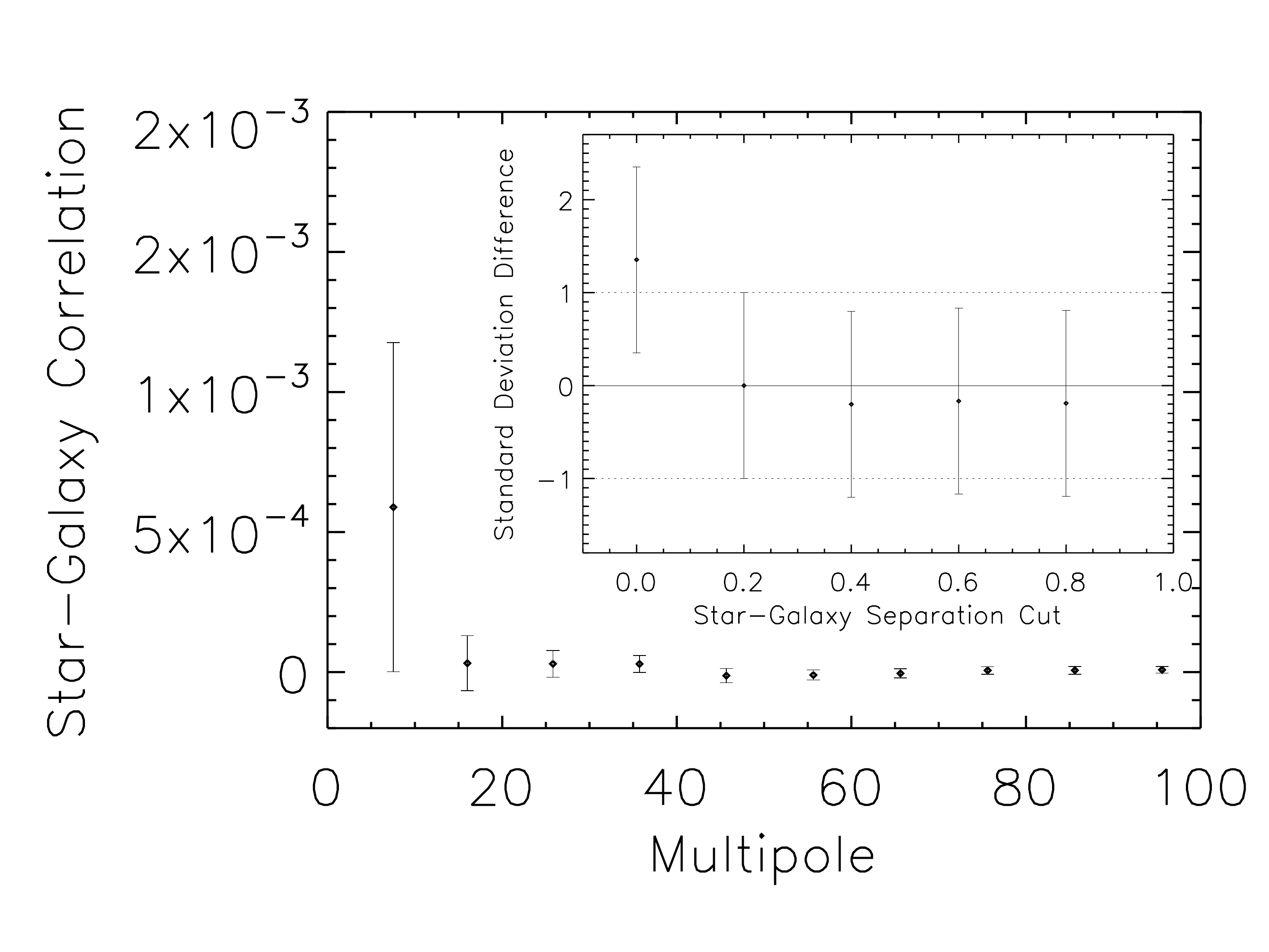}
      \includegraphics[width=8.4cm,height=8.4cm]{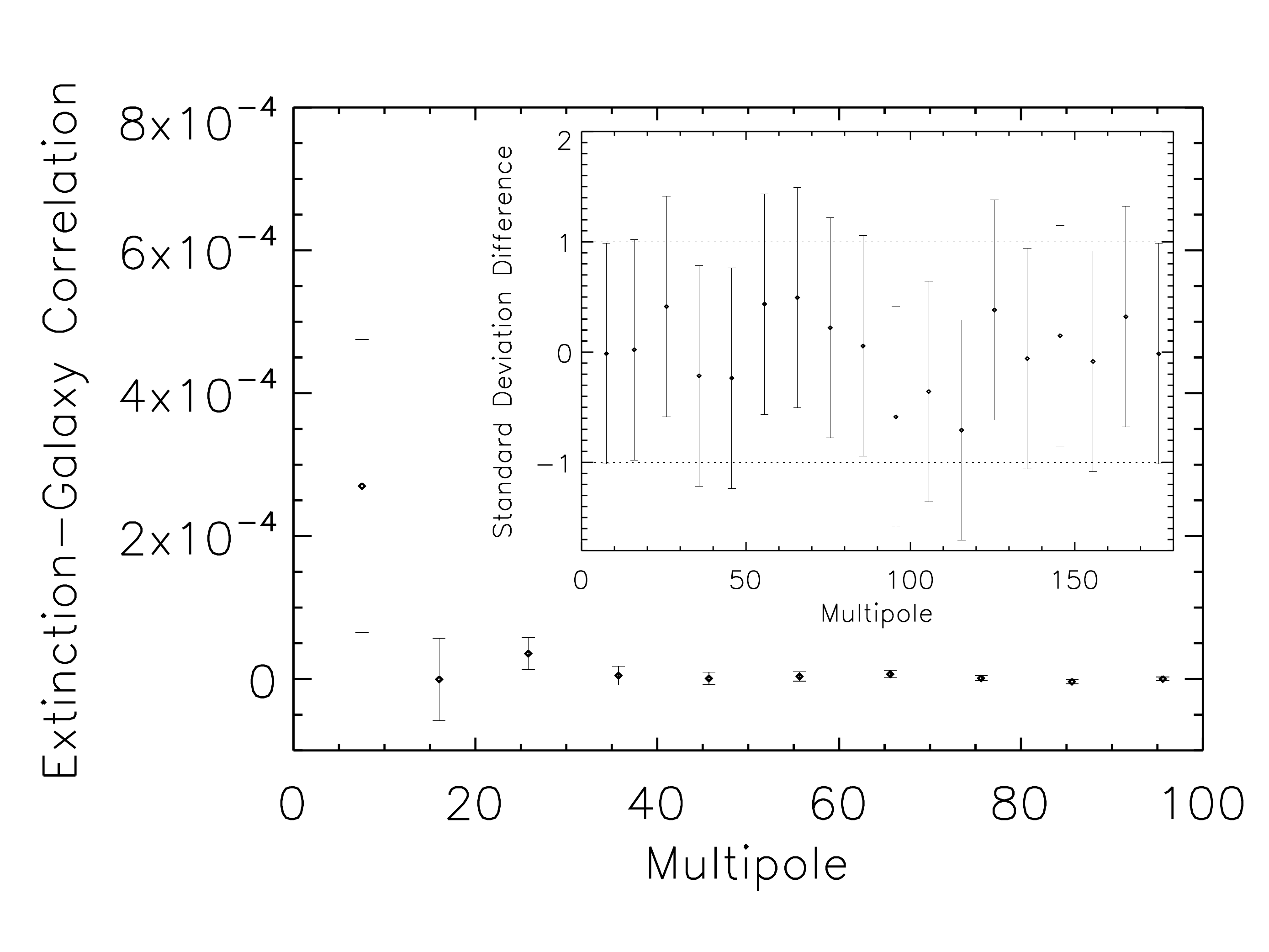}
    \end{minipage}
    \caption{\small{{\it Left Panel:} This is the cross correlation between galaxies in the highest redshift bin ($0.6 < z < 0.65$) and M-stars (main plot). No signal is observed over the profile and even the lowest multipole band is consistent with $1\sigma$. In addition, more severe star-galaxy separation cuts are found to give no change in the $C_{\ell}$ relative to the standard $\delta_{sg}>0.2$ cut (inset). The cuts correspond to remaining M-star contaminations of $5\%$ ($\delta_{sg}>0$), $1.5\%$ ($\delta_{sg}>0.2$), $1.2\%$ ($\delta_{sg}>0.4$), $0.8\%$ ($\delta_{sg}>0.6$) and $0.5\%$ ($\delta_{sg}>0.8$). {\it Right Panel:} The cross correlation between the same galaxies and Galactic extinction is again mostly inconclusive (main plot). For a further test we removed surveyed regions corresponding to high extinction ($> 0.1$ mag; $15\%$ of the survey area) and found no change to the measured spectra over the largest scales (inset).  }}
    \label{fig:stargalaxycorrelationplot}
  \end{flushleft}
\end{figure*}

{\it Star-Galaxy Separation} -- We selected the LRGs from the photometric sample using the selection criteria described thoroughly in \citep{Cannon06,Collister07,Blake07,Thomas10}. This was shown to be successful except for a $5\%$ M-star contamination. We acted to remove these objects, which vary across the Galactic plane, by imposing a cut on the star-galaxy separation parameter from ANNz ($\delta_{sg}>0.2$). As described in \citep{Collister07} this ensures that the contamination is minimised without losing too many real galaxies (quantified by comparing with the spectroscopic 2SLAQ sample). However, even with this cut the furthest redshift bin might still be affected given that it contains fewer natural objects. To test this we measure the cross correlation between M-star objects removed from the catalogue ($\delta_{sg} \le 0.2$) and galaxies within the range $0.6 \le z \le 0.65$. This is shown in the left panel of Figure~\ref{fig:stargalaxycorrelationplot}. 

We assign error bars using a generalisation of Equation~\ref{eq:gaussianerror} with the measured star and galaxy auto power spectra $C_{\ell}^{s}$ and $C_{\ell}^{g}$ as inputs. We also derive errors by cross correlating the aforementioned Gaussian simulations with the M-star distribution with no change in the conclusion: Overall there is no correlation between the two samples with the lowest multipole band consistent at $1\sigma$. Even though the correlation is not significant we go further and repeat the whole galaxy clustering measurement with a series of more aggressive ($\delta_{sg} \gg0.2$) star-galaxy separation cuts. The difference $C_{\ell}(\delta_{sg} =0.2) - C_{\ell}(\delta_{sg} = x)$ for the most anomalous band is illustrated as a function of star-galaxy cut in the inset of Figure~\ref{fig:stargalaxycorrelationplot}. We find {\it no} change over the scales of interest, implying minimal contribution from stars. The plot does highlight the importance of the initial cut however given the change when $\delta_{sg} = 0$.

{\it Extinction} -- Regions of high Galactic extinction could cause galaxies to be scattered from the sample as a function of sky position. The colour and magnitude cuts made on the photometric catalogue have been performed using extinction corrected model magnitudes. However, it could be that they contain errors that propagate into the analysis. With this in mind we therefore measure the cross correlation of galaxies and the extinction field. Similarly no clear signal is detected as seen in the right panel of Figure~\ref{fig:stargalaxycorrelationplot}. The lowest multipole band appears slightly suggestive but is not significant given the increased uncertainty from cosmic variance. To test this further we remove regions in the selection function corresponding to high extinction ($> 0.1$ mag) and repeat our analysis. This removes $\approx 15\%$ of the survey area and regions concentrated mainly at the edges of the survey geometry. We find this has a negligible effect on the overall $C_{\ell}$ profiles, including the largest scales, as seen in the same figure. 

We also repeated the clustering measurement analysis with the DR6 redshift catalogues from \citep{Abdalla08}. These galaxies represent only a $1\%$ reduction in area but have redshifts derived using a variety of template based procedures. This allows us to examine the extrapolation of the ANNz-2SLAQ training set with sky position. This is because the training set is limited to a narrow stripe within a limited region of the sky (and therefore extinction, for example) whereas the template based procedures are effectively blind to this calibration issue. We find no changes over large scales, which is consistent with the preliminary redshift tests performed in \citep{Abdalla08}.

Other systematics could remain and contribute to the $C_{\ell}$ measurement from an improper estimation of the selection function. Examples include variations in seeing, photometric calibration, over-lapping survey stripes and regions of low Galactic latitude. However, the previous MegaZ \citep{Blake07} and photometric study of \citep{Padmanabhan07} tested their profiles against these aforementioned effects and found no significant change across any scale.

{\it Alternative Models} -- Although more speculative a number of physical theories could produce a signature similar to our observed feature. For example, a modification to gravity or dark energy clustering \citep{Takada06} would give rise to an amplified signal over large scales. In particular changes to gravity would be enhanced within the statistic through redshift space distortions, which act to alter multipoles $\ell<50$ in the $C_{\ell}$ \citep{Thomas10,Fisher94,Padmanabhan07}. I.e., the distortions are sensitive to changes in $\ud {\rm ln} \, \delta / \ud {\rm ln} \, a$. Moreover, it would be interesting to see whether a complete non-linear treatment of redshift space distortions could influence these multipoles further. Some studies have also argued that large scale inhomogeneity or voids can give rise to the observed accelerated expansion. One would expect some alteration to the $C_{\ell}$ when the scales probed impinge upon any transition. However, the analysis would have to be altered for a comparison to that framework. In addition, significant non-Gaussianity from an exotic inflationary scenario is capable of causing an increase in biasing over large scales. Indeed, photometric surveys may prove to be one of the best methods to constrain non-Gaussianity in this way \citep{Dalal08}.

Naturally, any proposed explanation for this effect must be consistent with other probes and data. For example, alterations to the growth over large scales must be consistent with the CMB through the ISW effect. However, it is interesting that there are debates regarding various anomalies there too \citep{Copi10,Pontzen10}. Likewise, some analyses already imply tension with void-like models for acceleration \citep{Moss10}. Finally, it would be intriguing to see if the large amplitude implied in the power spectrum is similar to that which can produce bulk flows in velocity surveys \citep{Watkins09}.

{\it Conclusions} -- Using the largest ever galaxy survey we find an excess of clustering in the angular power spectra of Luminous Red Galaxies (LRGs) that is not predicted by standard cosmology. This is evident over low multipoles/large scales, particularly in the highest redshift bin. This could be the consequence of systematics that affect the furthest, faintest and photometrically least reliable galaxies. Alternatively, this could be the sign of new physics over the largest scales probed by the survey, which would occur at the highest redshift.

We tested the survey, statistics and quantified the anomalous feature. We found no evidence to suggest it was caused by systematics from Galactic extinction, survey geometry, star contamination or the redshift estimation method. Excess power was measured in all four bins but was $\sim 4\sigma$ at the highest redshift ($0.6 < z < 0.65$). Furthermore, any boost to the power spectrum emerges at an effective scale $k \lesssim 0.01 \mathrm{h \, Mpc^{-1}}$ or $\lambda \gtrsim 700 \, \mathrm{h^{-1} \, Mpc}$. We concluded by highlighting a number of theories that could give rise to this effect, such as exotic dark energy, modified gravity, changes to redshift space distortions, large scale inhomogeneity or non-Gaussianity.

It will be fascinating to see if large volume spectroscopic surveys, such as BOSS, observe such anomalies in the future given the different dependence on systematics. However, one might expect methodologically similar photometric surveys like the Dark Energy Survey to observe these features again. Clearly, even more effort to understand factors affecting completeness or to estimate a non-discrete selection function will be invaluable.

{\it Acknowledgements} -- ST acknowledges UCL's Institute of Origins for a Post-doctoral Fellowship. FBA and OL acknowledge the support of the Royal Society via a Royal Society URF and a Royal Society Wolfson Research Merit Award, respectively. We acknowledge use of Healpix \citep{Gorski05}. Thanks to Hume Feldman, Gert Huetsi, Hiranya Peiris, Sarah Bridle and Jochen Weller for useful discussions.

\bibliography{paper}

\end{document}